\documentstyle[12pt]{article}
\newtheorem{main}{Proposition}[section]
\def\real{{\rm I\kern-.2em R}}
\def\complex{\kern.1em{\raise.47ex\hbox{
            $\scriptscriptstyle |$}}\kern-.40em{\rm C}}
\title{Odd Viscosity }
\author{J.~E.~Avron}
\author{J.~E.~Avron\\ Department of Physics, Technion, 32000 Haifa, Israel }

\begin{document}
\maketitle 
\begin{abstract} When time reversal is broken  the viscosity
tensor can have a non vanishing odd part. In two dimensions, and only then, such
odd viscosity is compatible with isotropy.  Elementary and basic features of
odd viscosity are examined by considering  solutions of the wave and
Navier-Stokes equations for hypothetical fluids where the stress is
dominated by odd viscosity.

\end{abstract}

\section{Introduction and Overview}
Normally, one associates viscosity with dissipation. However, as the viscosity
is, in general, a tensor, this need not  be the case since  the
antisymmetric part of a tensor is not associated with
dissipation. We call the antisymmetric part  odd. It must
vanish, by Onsager relation, if time reversal holds. It must also vanish
in three dimensions if the tensor is isotropic. But, in two dimensions
odd viscosity is compatible with isotropy. 

It is conceivable that that  odd viscosity does not vanish for many system
where time reversal is broken either spontaneously or by external fields.  But, 
I know of only two systems for which there are  theoretical studies of the odd
viscosity and none for which it has been studied experimentally. In
superfluid  He$^3$,  time reversal and isotropy can  spontaneously
break and the odd viscosity has three independent components
\cite{VW}. As far as I know there is no estimate for their magnitudes. In the two
dimensional quantum Hall fluid time reversal is broken by an external
magnetic field. In the case of a full Landau level the dissipative viscosity
vanishes.  The odd viscosity,
$\eta^a$, has been calculated for non interacting electrons in
\cite{asz} for the lowest Landau level. Using results of
\cite{levay}, the odd viscosity for any integral
filling factor
$n$ is:
\begin{equation}
{\eta^a}(n) =\frac{n^2}{2}\, \frac{eB}{4c},\quad n=1,\dots.\label{one}
\end{equation}
$e$ is  the charge of the electron, $c$ the velocity of light.
 It is an amusing coincidence of the cgs units that
the  {\em kinematic viscosity} of the electron gas, at integer fillings,
\begin{equation}
\left(\frac{\eta^a}{\rho_e}\right)(n)= n\,\frac{h}{8m_e},\label{kv}
\end{equation}
is close to $n$ in cgs units.

 Significant odd viscosity can be responsible for odd properties and
we shall  illustrate these by considering the wave and Navier-Stokes equations
for hypothetical media where the odd viscosity dominates.
 A good example for such a peculiar property is the
following:  Consider a small, slowly rotating circular cylinder (so that the
Reynolds number is small) in an fluid which has both dissipating and and odd
viscosity. The dissipative viscosity applies a torque which resists the
rotation. This agrees with common intuition. The odd viscosity leads to {\em
radial}  pressure on the cylinder which is proportional to  the rate of rotation
(and the coefficient of odd viscosity, of course). Reversing the orientation of
rotation, reverses the sign of this pressure. This response to the rotation is
not particularly intuitive.

Materials with significant odd viscosity, be they solids or liquids, can support
non-dissipating chiral viscosity waves with a quadratic dispersion. 
The reflection from a boundary of these  waves obeys a different rule
than the reflection of acoustic waves. In addition, such materials can function
as circular polarizers for ordinary acoustic waves. At the same time,
viscosity waves can be quite elusive. In particular we shall show that there are
no viscosity waves in isotropic and incompressible media. In particular, the
Hall system, being  isotropic and incompressible,  does not support viscosity
waves. 

A dimensioless number that gives a measure of the importance of
viscosity relative to  elasticity is
\begin{equation}
\varepsilon =\frac{\nu \omega}{c^2_s}
\end{equation} where $\nu$ is the kinematic viscosity, $\omega$ the frequency,
and $c_s$ the velocity of sound. Odd viscosity is always
unimportant relative to elasticity at low frequencies.  For the dissipating
viscosity, and also for the odd viscosity in the Hall effect, the kinematic
viscosity  is of order
$1$ in cgs units. Then, since
$c_s$ is typically of order
$10^{5}$ in cgs, $\varepsilon$ is of order unity for $\omega$ of order GHz.

As we shall see, the generalized Navier-Stokes equation that allows for odd
viscosity preserves the basic properties of fluid dynamics of the ordinary
Navier-Stokes equation:
Kelvin theorem and Bernoulli law generalize to non zero odd viscosity.

\section{Odd Viscosity}
Consider  a hypothetical,
homogeneous  and ideal Newtonian fluid. The stress due to
viscosity is: 
\begin{equation}
\sigma^v_{ij}=\eta_{ijkl}\dot u_{kl} .\label{visco}
\end{equation}
where $u_{ij}=\frac{1}{2}\left(u_{i,j}+u_{j,i}\right)$ is the strain,
$\dot u_{jk}$ is the strain
rate, $\eta $ a (constant) viscosity tensor and $\sigma$ the stress.  By general
principles,
\cite{llp}, the viscosity tensor,
$
\eta_{ijkl}$, is symmetric under 
 $i\leftrightarrow j$, and  $k\leftrightarrow l$. One can always write $\eta =
\eta^S+\eta^A$ where
$\eta^A $ is anti symmetric under  $\{ij\}\leftrightarrow \{kl\}$ and $\eta^S$ is
symmetric under  $\{ij\}\leftrightarrow \{kl\}$.
By Onsager relation \cite{onsager}  the anti-symmetric part is odd under time
reversal and the symmetric part is even, e.g., with $B$ an external magnetic
field;
\begin{equation}
\eta^S(B)=\eta^S(-B), \quad \eta^A(B)=-\eta^A(-B).
\end{equation}
So, for $\eta^A\neq 0$ time reversal must be broken.
One can ask if odd viscosity is not just a way of mascarading, say, an external
magnetic field. The answer to this is no. This can be seen by a counting
argument.  A (constant) magnetic field has three components while the space of
anti-symmetric tensors 
$\eta^A$ is 15 dimensional in three dimensions.

\subsection{Viscosity in Two Dimensions} 
In two dimensions there is a natural basis for representing real 4-th rank tensor
which are symmetric in pairs\footnote{This basis was suggested by
the referee of this paper.}:
\begin{equation}
\eta=\sum \eta_{ab}\  \sigma^a\otimes \sigma^b, \quad a,b\in \{0,1,3\}.  
\end{equation}
Here $\sigma^a$ are the Pauli matrices with $\sigma^0$ the unit matrix. 
$\sigma^2$, which is conventionally imaginary, is not used. In components:
\begin{equation}
\eta_{ijkl}=\sum \eta_{ab} \sigma^a_{ij} \sigma^b_{kl} \quad i,j\in \{0,1,3\}.  
\end{equation}
The space of such tensors is nine dimensional, and its splits to a six
dimensional even part and three dimensional odd part. The odd part takes the
form
\begin{equation}
\eta^A=\sum_{a\neq b} \eta^A_{ab}\ (\sigma^a\otimes \sigma^b-\sigma^a\otimes
\sigma^b)
\quad i,j\in
\{0,1,3\}.  
\end{equation}
To see what happens in isotropic media, recall that $i\sigma^2$ is the
generator of rotations in two dimension, and it anticommutes with $\sigma^a$
for $a=1,3$ and commutes for $a=0$. Hence
$\sigma^2\otimes\sigma^2$ commutes with $\sigma^a\otimes\sigma^b$ if $a,b=1,3$
and
$a=b=0$. The isotropic symmetric part is two dimensional and is
characterized by two viscosity coefficients:
\begin{equation}
\eta^S={\eta^s}\ (\sigma^1\otimes \sigma^3+\sigma^3\otimes
\sigma^1)+\zeta\ \sigma^0\otimes \sigma^0.  
\end{equation}
The odd isotropic part is one dimensional and is of the form
\begin{equation}
\eta^A={\eta^a}\ (\sigma^1\otimes \sigma^3-\sigma^3\otimes
\sigma^1).  
\end{equation}
The non-zero components are determined by
\begin{equation}
\eta^A_{1122}=0, \ \eta^A_{1211}=-\eta^A_{1222}=\eta^a .
\end{equation} 
It is sufficient that a two dimensional medium is invariant under rotation by
$\pi/4$ for $\eta$ to have this form.

As a consequence of this, isotropic two dimensional media are characterized, in
general, by {\em three} coefficients of viscosity: two for the even part
and the third for the odd viscosity $\eta^a$.
This is in contrast with a claim made in \cite{chorin} (p.\ 45-46), and in
\cite{ll} $\S 15$, that isotropy alone implies that the viscosity tensor has two
coefficients of viscosity.
 
The stress $\sigma^v$ associated with the viscosity of an isotropic medium, is 
\begin{eqnarray}
\sigma^v_{ij}&=& \eta^s\,(\dot u_{ij}+\dot
u_{ji})+\zeta^s\delta_{ij}\dot u_{kk}\nonumber\\ 
&&-2\eta^a(\delta_{i1}\delta_{j1}-\delta_{i2}\delta_{j2})\dot u_{12}+\eta^a
(\delta_{i1}\delta_{j2}+\delta_{i2}\delta_{j1})(\dot u_{11}-\dot u_{22}).
\end{eqnarray} 
In  an  incompressible fluid $\dot u_{kk}=0$ and the stress
becomes independent of $\zeta$, the second viscosity coefficient.
Because of this 
$\zeta$ plays no role in the Navier Stokes equation for incompressible
fluids.  Incompressible and isotropic fluids in two dimensions with broken time
reversal are  characterized by two viscosity coefficients one for the
odd part,
$\eta^a$, and one for the even part $\eta^s$.

\subsection{Viscosity in Three Dimensions} In three dimensions isotropy  implies
that
$\eta^A=0$. Three dimensional isotropic media are characterized by only
{\em two} coefficients of viscosity, and both of these are associated with
the even part of the viscosity tensor. Isotropic and incompresible media are
characterized by a single dissipative viscosity coefficient. 

The vanishing of the odd viscosity for isotropic tensors can be seen by the
following elementary argument, which I owe to L. Sadun.  Vectors in three
dimensions are associated with the
${\bf 1}$ representation of the rotation group. 2-tensors are identified with 
$${\bf 1}\otimes {\bf 1} = {\bf 2}\oplus {\bf 1}\oplus {\bf 0}
$$ representation of the rotation group. The vector representation ${\bf 1}$ on
the right hand side, is associated with pseudo
vectors and so with the antisymmetric 2-tensors.  The 
${\bf 2}\oplus {\bf 0}$ is the six dimensional representation associated with the
symmetric 2-tensors.  This shows that there is one isotropic 2-tensor
in three dimensions, namely the identity. Continuing in this vein,  tensors
$t_{ijkl}$ with the symmetry 
$i\leftrightarrow j$, and  $k\leftrightarrow l$ are identified with
\begin{equation}({\bf 2}\oplus {\bf 0})^2={\bf
4}\oplus{\bf 3}
\oplus3\cdot{\bf 2 }\oplus{\bf
1}\oplus2\cdot{\bf 0}.
\end{equation}
This shows that the space of isotropic 4-tensors with the symmetry 
$i\leftrightarrow j$, and  $k\leftrightarrow l$  is 2-dimensional.
  The isotropic
4-tensors $t^S$ make a two dimensional family, given by
\begin{equation}
t^S_{ijkl}= t_1 \delta_{ij}\delta_{kl} + t_2
\left(\delta_{ik}\delta_{jl}+\delta_{il}\delta_{jk}\right),\label{elasticiso}
\end{equation}
this leaves nothing for
the isotropic
antisymmetric 4-tensor  and so $\eta^A=0$.

The number of independent components of the odd and even part of the viscosity
tensor is given in the table.

\begin{center} 
\begin{tabular}{|c|c|c|c|c|}\hline
Dimension&
           \multicolumn{2}{c|}{Even}&\multicolumn{2}{c|}{Odd} \\ \cline{2-5} 
& general&isotropic& general&isotropic \\ \hline
2&9&2&3&1 \\ \hline
3&21&2&15&0 \\ \hline
\end{tabular}
\end{center}

\section{Viscosity Waves} 
Consider a hypothetical ideal
three dimensional   medium where the stress tensor is dominated by
the odd viscosity.  For small oscillations the equation of motion is 
\cite{llp}:
\begin{equation}
\rho \ddot{u}_i= \partial_j \sigma_{ij}^v.\label{newton}
\end{equation}
where $\rho$ is the (mass) density, $u_j$ the (cartesian) j-th component of the
displacement (assumed small), and
$\sigma^v$, the stress tensor, depends linearly on the strain rate.  Linear
elasticity is  mathematically trivial in the sense that solving the PDE reduces
to a problem in linear algebra. In a
homogeneous systems 
$\rho$ is a constant so one can, without loss,
set
$\rho=1$ (by absorbing
$\rho$ in $\sigma$.)  
For a medium with odd viscosity Newton equation gives a linear PDE which is
second order in space and time:
\begin{equation}
\ddot u_i =\eta^A_{ijkl}
\dot u_{l,jk}.\label{vw}
\end{equation}
This equation is {\em not} just an alternative
way of describing the Lorentz force. The corresponding  Newton-Lorentz equation:
\begin{equation}
\ddot {\bf u}+ {\bf  B}\wedge \dot {\bf u}  =0,
\end{equation}
 is not a PDE.

Newton equation for viscosity waves admits an integral and by choosing the
constant of integration appropriately, reads:
\begin{equation}
\dot u_i =\eta^A_{ijkl}
u_{l,jk}.\label{vw}
\end{equation}
This equation is first order in time and second order in space. Because
$\eta^A$ is antisymmetric it has the same character as Schr\"odinger
equation\footnote{In real coordinates, with $\psi_1=Re\,\psi$ and
$\psi_2=Im\,\psi$, Schr\"odinger equation for a free particle takes the form
$\dot\psi_i=\epsilon_{ij}\psi_{j,kk}$.}. But it is a classical equation in the
sense that the wave function is directly observable. 

Consider plane waves solutions to
this equation with wave vector
${\bf k}$. Let
${\bf \Omega(k)}$ be the (pseudo) vector with components that are quadratic
forms in the coordinates $k_j$:
\begin{equation} \Omega_i({\bf k})=\epsilon_{ijk}\eta^A_{j\alpha
\beta k}k_\alpha k_\beta.\end{equation} Newton law for a plane wave propagating
in the
${\bf k}$ direction can be written as:
\begin{eqnarray}
\dot {\bf u} =
 {\bf \Omega}({\bf k})\wedge {\bf u}.\label{dispersionv}
\end{eqnarray}
From this equation it is clear that the viscosity wave in three dimensions is
chiral and circularly polarized in the plane perpendicular to the vector
${\bf \Omega}({\bf k})$ and has quadratic dispersion.
Recall that ordinary sounds waves
have linear dispersion, but, rods and plates also allow for elastic modes
with quadratic dispersion \cite{llp}. The equation for viscosity waves
equation of motion, Eq.~(\ref{dispersionv}), is formally identical to the Landau
Lifshitz equation for magnons
\footnote{I thank Dr. M. Milovanovic for reminding me of
this fact.}.

\subsection{Incompressible Media} Consider an incompressible medium where  ${\bf
\nabla \cdot 
 u}=0$. For a plane wave this means that ${\bf \Omega}$ is parallel to ${\bf
k}$, that is:
\begin{equation}{\bf \Omega}\wedge {\bf k}=0, \Leftrightarrow 0=\eta^A_{ijkl}
k_jk_kk_l. 
\end{equation}  
For a viscosity wave propagating in the $1$ direction 
\begin{equation}{\bf\Omega({ k})}= k^2 \, \pmatrix{
\eta^A_{2113}\cr \eta^A_{3111}\cr \eta^A_{1112}\cr}.\end{equation}
From this it follows that  in an incompressible medium  either
$\eta^A_{i111}=0$ for all
$i$ or ${\bf u}=0$ identically. This leads to:
\begin{main} There are no odd viscosity waves in 
isotropic and incompressible fluids. In particular, there are no viscosity
waves in a two dimensional quantum Hall fluid.
\end{main}
This is so because  in two dimensions   isotropic fluids the non zero
component of the isotropic odd viscosity are
$\eta^a=-\eta^A_{1222}=\eta^A_{2111}$. For an isotropic fluid we can choose the
$1$ axis to coincide with the wave vector, and then incompressibility says
that either $\eta^a=\eta^A_{2111}=0$ or the wave has zero amplitude. 
\footnote{The result, and proof, works also for a two dimensional planar medium
embedded in three dimensions.} 
In three dimensions, there are no viscosity waves because the odd viscosity
vanishes.

\subsection{The Energy Flux of Viscosity Waves} 
 In the case $\eta^S=0$ one can define a
conserved current associated with conservation of energy. The kinetic energy
density:
\begin{equation}
{\cal E}= \frac{1}{2}\left(Re\, {\bf \dot{ u}}\right)^2,
\end{equation}
satisfies a conservation law:
\begin{equation}
\partial_t{\cal E}= \partial_iJ_i,
\end{equation}
where the energy flux $J$ is
\begin{equation}
J_i=\eta^A_{ijkl}\, Re \,( \dot u_j) \, Re (\dot u_{kl})= \sigma^v_{ij} Re(\dot
u_j).
\end{equation}
It follows that odd viscosity carries energy flux. There is a generalization
of this energy flux to media where the stress also has elastic part.

\subsection{Dissipative Shear Waves}   It is instructive to contrast
viscosity waves with the strongly dissipative shear waves \cite{ll}.  For shear
waves 
$\eta^S\neq 0$ and
$\eta^A=0$, and the dispersion curves 
 are {\em pure imaginary}
\begin{equation} i \omega=-\eta^S k^2,\quad \eta^S,\omega>0\quad Im \, k>0.
\end{equation}  
Shear waves decay  rapidly, by a factor of about
$\exp (2\pi)\approx 540$ in one period \cite{llp}. If one now consider a mixed
situation where $\eta^S$ and $\eta^A$ are of the same magnitude  then the
attenuation is much weaker---of order $\exp (2\pi \tan(\pi/8))\approx 13$.

\section{Scattering of Viscosity Waves} Using the rules of
geometric acoustics it is straightforward to find the laws of reflection and
refraction of an acoustic wave (in a medium where $\eta =0$) from an interface
with an odd viscous medium (where the stress is dominated by $\eta^A$). If we let
$y$ be the axis separating the two media, then $k_2$ is conserved and so is the
frequency
$\omega$. Let
$\theta$ be the angle of incidence, $\cos (\theta)=\hat x\cdot \hat k_i$,  of an
acoustic wave with sound velocity $c_s$ and
$\phi$ the angle of the transmitted viscosity wave, 
$\cos (\phi)=\hat x\cdot\hat k_t$ .  Then a little geometry shows that
\begin{equation}
\sin^2\phi = \varepsilon \sin^2 \theta,\quad \varepsilon= \frac{\eta
\omega}{\rho c_s^2}.
\end{equation}
 We see that $\varepsilon^{-1/2}$ plays the role of the index of reflection. 
Since  $\varepsilon$  is small at low frequencies, the index of
reflection is large, and the transmitted beam is essentially normal to the
interface. Since the index depends  on
$\omega$ the reflection is dispersive. 

\subsection{Reflection from Empty Space} Consider the reflection of ideal
viscosity waves in {\em two dimensional isotropic and  compressible}
medium from an interface with a vacuum which we take to be the line $x=0$. 
 A  viscosity waves is
\begin{equation} {\bf u}(z) = \pmatrix{1\cr i\cr}\, z,\quad z=\exp i({\bf
k}\cdot {\bf  x} -\omega t),
\end{equation}
where $\omega =\eta^a k^2$. The waves is circularly polarized in the plane and 
changes polarization from longitudinal to transversal.
The
boundary conditions at $x=0$ are 
\begin{equation}
\sigma_{11}=\sigma_{12}=0.\end{equation} For this wave:  
\begin{equation}
\sigma_{11}({\bf u}) = -\eta_{1112} \,\bar k,\quad \sigma_{12}({\bf u}) =
-i\eta_{1112}\, \bar k; \quad k=k_1+ik_2, \quad \bar k=k_1-ik_2.\label{st}
\end{equation} 
Let
$k=(k_1,k_2)$ be the incident wave,  $\tilde k = (-k_1,k_2)$ be the reflected
wave;
 and $\tilde z=\exp i(\tilde {\bf k}\cdot {\bf x}
-\omega t)$. ${\bf  u}(z)$ is the incident wave (from the left) and ${\bf 
u}(\tilde z)$ the reflected wave and $R(k)$ the  reflection amplitude. 
The wave on the left is:
\begin{equation}
{\bf  u}(z)+ R(k)\,{\bf  u}(\tilde z),\quad x<0.
\end{equation}
Substituting in the boundary conditions, using Eq.~(\ref{st}) and linearity, one
finds 
\begin{equation}R(k)=\frac{k}{\bar k}.\end{equation} The phase of
the reflected wave has an interesting dependence on the direction of incidence:
It goes from
$1$ for normal reflection to
$-1$ for grazing reflection. This is quite unlike the reflection of say
longitudinal sound waves in an incompressible liquid  where
the reflection amplitude is independent of the direction of incidence.

\subsection{Odd Viscosity as Polarizer}  Consider a wave propagating in the
$x$ direction, of a three dimensional incompressible medium, so that to the left
there is an isotropic elastic medium with velocities of sound $c_t$ and
$c_\ell$. To the right is a viscous (possibly incompressible) medium.
 The $y-z$
plane separates the two media. Incompressibility says that 
${\bf \Omega(k)}$ points in the $x$ direction.
Let 
\begin{equation} {\bf e}_+=(0,1,i), \quad {\bf e}_-=(0,1,-i),\end{equation}
denote the two basic vectors of circular polarization. Consider scattering of a
transverse wave with positive chirality from the $y-z$ interface. The incoming
and reflected waves in the elastic medium are
\begin{equation}
{\bf e}_+\Big( \exp ik(x-c_t t) +r\, \exp ik(-x-c_t t)\Big), \quad x<0.
\end{equation}
The transmitted wave (necessarily with positive chirality) is: 
\begin{equation}
t\,{\bf e}_+ \exp i(\tilde k x-\omega t),\quad x>0 .
\end{equation}
Where $\omega = k c_t =|{\bf \Omega(\tilde k)}|$.
The basic boundary conditions  matches $\sigma_{1j}^L$ on the left with
$\sigma_{1j}^R$ on the right.
One checks that
$\sigma_{11}^L= \sigma_{11}^R=0$ and that 
$
i\sigma_{12}^R =\sigma_{13}^R.
$
 $i\sigma_{12}^L =\sigma_{13}^L$ is automatically satisfied ( confirming the
anzatz that there is no  reflected wave with flipped chirality). The remaining
equation,  $\sigma_{12}^L =\sigma_{12}^R$, gives
\begin{equation}
1-r=\varepsilon t, \quad \varepsilon=\frac{\eta_{2113}\omega}{\rho c_t^2}>0.
\end{equation}
Continuity of the wave $u$ on the boundary gives two equations but
only one new: $1+r=t$.  From these I finally get:
\begin{equation}
t=\frac{2}{1+\varepsilon},\quad r=\frac{1-\varepsilon}{1+\varepsilon},\quad
\varepsilon=\frac{\eta_{2113}\omega}{\rho c_t^2}.
\end{equation}
 Note that $|r|\le 1$ as it
should: The  reflected wave always has smaller amplitude than the incident
wave.  However, for $t$ one gets an unusual behavior: $0\le t\le 2$ and the
amplitude 
$t$ is {\em maximal} for $\varepsilon =0$ when the {\em amplitude} of the
transmitted wave is twice as large as that of the incident wave! This
surprise is mitigated when one notices that for the energy flux  in the $x$
direction,
$J_1$, the dependence on
$\varepsilon$ is  $\frac{\varepsilon}{(1+\varepsilon)^2}$, which gives
maximal flux at $\varepsilon=1$ as one expects, since in this case $r=0$. The
wave is perfectly transmitted. 

Consider now the scattering of a viscosity wave so that the incident wave has
{\em negative} chirality.
 The most general transverse wave in the elastic medium with the given incident
wave
  is:
\begin{equation}
r{\bf e}_+\, \exp ik(-x-c_t t) + {\bf e}_-\Big( \exp ik(x-c_t t) +\tilde r\,\exp
ik(-x-c_t t)\Big),
\end{equation}
while the wave in the viscous medium is as before. Matching 
$\sigma_{1j}^L$ with $\sigma_{1j}^R$  one finds that as before
$\sigma_{11}^L= \sigma_{11}^R=0$ and $
i\sigma_{12}^R =\sigma_{13}^R.
$
Writing out $i\sigma_{12}^L =\sigma_{13}^L$ gives:
 $\tilde r=1$.
The remaining equation,  $\sigma_{12}^L =\sigma_{12}^R$, gives
\begin{equation}
r=\varepsilon t.
\end{equation}
This is all that follows from the continuity of the three stress
components. Imposing  continuity of the wave on the boundary  gives
{\em two} new equations:
\begin{equation}
r=t,\quad 1+\tilde r=0.
\end{equation}
The second equation is in conflict with the equation for the continuity of the
stress\footnote{This is the generic situation
in scattering of acoustic and viscosity waves. That is, in general imposing
continuity of the wave across the interface gives an overdetermined system.
In this sense, the previous example is exceptional in that both the stress
and the wave turned out to be continuous.}. The way out is {\em not} to require
continuity of the wave, but instead continuity of the energy flux. This  holds
if the flux vanishes at the surface  and  sets
$r=t=0$. The incident wave with the wrong polarization is totally reflected.  
We see that  {\em anisotropic incompressible  media with odd viscosity  act as 
circular polarizers.}

\section{Navier Stokes Equation} Consider the Navier Stokes
equation, for homogeneous ($\rho =1$),  isotropic and incompressible fluid in two
dimensions--the one dimension where odd viscosity is compatible with
isotropy. 

With broken time reversal, the general Navier-Stokes equation for the pressure
$p$ and the velocity field ${\bf v}$,  is the obvious generalization of the
standard Navier-Stokes equation \cite{chorin}:
\begin{equation}
\partial_t{\bf v}+({\bf v}\cdot \nabla){\bf v} =-\nabla p +
\eta^s\Delta {\bf v} - \eta^a \Delta {\bf v}^*;\quad
\nabla\cdot {\bf v}=0.\end{equation}
The dual is defined, as usual, by
\begin{equation}
{ v}_i^*= \epsilon_{ij}{ v}_j.\end{equation}
Incompressibility is  implemented by introducing a stream function and the
equations can be  alternatively  written as four equations for the four fields
$(p,\psi,{\bf v})$:
\begin{equation}
\partial_t{\bf v}+({\bf v}\cdot \nabla){\bf v} =-\nabla (p -\eta^a\xi)+
\eta^s\Delta {\bf v};\quad
 {\bf v}=(\nabla \psi)^*,\quad \xi=-\Delta\psi
.\end{equation} Taking curl the equation for the
vorticity is
\begin{equation}
\partial_t\xi + ({\bf v}\cdot\nabla)\,\xi = 
\eta^s\Delta {\bf \xi} .
\end{equation}
Vorticity is be generated only by the symmetric (dissipative)
part of the viscosity tensor. The odd viscosity does not generate 
vorticity.

\subsection{Bernoulli Law} Consider  incompressible fluid in two
dimensions, with $\eta^s=0$, in steady state. The equation of motion is:
\begin{equation}
({\bf v}\cdot \nabla){\bf v} =-\nabla (p -\eta^a\xi).\end{equation}
Since
\begin{equation}
\frac{1}{2}\, \nabla( {\bf v}^2)=  ({\bf v}\cdot \nabla){\bf v} +{\bf v}\times
(\nabla\times {\bf v})
=-\nabla (p -\eta^a\xi) +{\bf v}\times
(\nabla\times {\bf v}),
\end{equation}
one gets, by integrating along a streamline, that 
\begin{equation}
\frac{1}{2}\, {\bf v}^2+ p -\eta^a\xi=const.
\end{equation}
 This looks at first like an interesting generalization of Bernoulli's
law to odd viscosity with the pleasant feature that the vorticity 
(weighted by the odd viscosity) plays a role of (signed) kinetic energy.
However, as vorticity is conserved along a streamline, it is actually precisely
equivalent to Bernoulli's law.

\subsection{Stokes Equation}  The  limit of small Reynolds number is described
by Stokes equation. This is the case where normally viscosity dominates. Stokes
equation is linear, so this is also the easy limit. In the present context,
with broken time reversal, and for two dimensional isotropic and incompressible
fluid:
\begin{equation}
\partial_t{\bf v} =-\nabla (p -\eta^a \xi) +
\eta^s\Delta {\bf v};\quad
 {\bf v}=(\nabla \psi)^*.\end{equation}
Taking curl and div we get:
\begin{equation}
\partial_t\xi = 
\eta^s\Delta {\bf \xi},\quad 0=-\Delta (p -\eta^a \xi) .
\end{equation}
In a steady state, if $\eta^S\neq 0$, both the pressure and the vorticity must
be harmonic functions and $\psi$ bi-harmonic.

\subsection{A Rotating Disc} Consider a rigid disc which is rotating with
constant (and small) angular velocity $\Omega$ in a fluid subject to no-slip
boundary conditions. Stokes equation   can be solved explicitly
in this case \cite{shvil,shapere}.  I will show that while the ordinary
viscosity applies a torque that resists the rotation of the disc, the odd
viscosity leads to a pressure on the disc which is proportional to the angular
velocity. The pressure  can be either positive or negative,
depending on the sense of rotation.

Written in complex notation Stokes equation read:
\begin{equation} \bar \partial\left(p+i\eta \xi\right)=0, \quad
\eta=\eta^s+i\eta^a.
\end{equation}
Since $\psi$ is bi-harmonic and real
\begin{equation}
-2i \psi = z\, \overline{a(z)} - \bar z\, a(z) + b(z) -\overline{b(z)}.
\end{equation}
 Let $v=v_1+iv_2$.  Using
\begin{equation}
v=-2i \bar\partial \psi,\quad \xi=-2i \partial v.
\end{equation}one finds
\begin{equation}
v= -a(z) + z\, \overline{a'(z)} -\overline{b'(z)}, \quad\xi=2i
(\overline{a'(z)}-{a'(z)}).
\end{equation} 
Since the velocities at infinity are finite, one must have\footnote{One can
actually allow
$a_{-1}$ real. One can not allow a log term in $a(z)$ if one is interested
in continuous solutions for the velocity field.}
\begin{equation}
a(z)=  \sum_{j=0}^\infty\ \frac{a_j} {z^j},\quad b(z)= b\log z +
\sum_{j=-1}^\infty\ \frac{b_j} {z^j}.
\end{equation}
On the surface of the circle
 $\bar z\, z=1$ and  one can trade $\bar z$ for an inverse power of $z$, so:
\begin{equation}
i\Omega z=-\sum_{j=0}^\infty\ \frac{(j-1)\bar b_{j-1}+a_j} {z^j}
-\sum_{j=3}^\infty\ (j-2) \bar a_{j-2} z^j -\bar b z, \quad |z|=1.
\end{equation}
Equating coefficients we get
\begin{equation} i\Omega=-\bar b.\end{equation}
All other coefficients are zero. That is
\begin{equation}
v=- i\,\frac{\Omega}{\bar z},\quad \xi=0. \end{equation}
It then also follows that $p$ is a constant.
Using the velocity field one finds  the strain on the surface of the circle:
\begin{equation}
\dot u_{11}= -\Omega \sin (2\theta), \quad \dot u_{12}= \Omega \cos (2\theta),
\quad \dot u_{22}=\Omega \sin (2\theta).
\end{equation}
Here $x= |z| \cos\theta$.
Let us write the stress as $\sigma=\sigma^s+\sigma^a$ where the  first piece is
due to the dissipative viscosity  and the second is due to the odd
viscosity. Then
\begin{equation}
\sigma^s_{ij}=2\,\eta^s \dot u_{ij}.
\end{equation}
And
\begin{equation}
\sigma^a_{11}=2\eta^a \Omega\,\cos(2\theta),\quad 
\sigma^a_{12}=2\eta^a \Omega\,\sin(2\theta),\quad 
\sigma^a_{11}=-2\eta^a
\Omega\,\cos(2\theta).
\end{equation}
Let $n$ and $t$ denote the unit vectors normal and tangent to the disc. The a
calculations gives
\begin{equation}
\sigma^s_{nn}=2\eta^a\Omega,\quad 
\sigma^s_{tn}=2\eta^s \Omega.
\end{equation}
This says that the symmetric viscosity resists the rotation by applying a
torque on the rotating circle. The  odd viscosity applies no torque, but
instead, a normal pressure on the circle, which is proportional to the rate of
rotation.

{\it Remark:} One could have also asked what is the effect of odd viscosity
on  the drag and lift (the Magnus force). Unfortunately, for
an incompressible, viscous fluid in two dimensions the question is moot: Stokes
equation in two dimension does not admit steady state solutions that describe a
moving disc with no slip boundary conditions  in a fluid that is at rest at
infinity. Two dimensions behave like one dimension and unlike three dimensions.
This feature of the Stokes equation  is known in classical fluid mechanics as the
Whitehead paradox.

 \section*{Acknowledgment}
I thank Prof. Mikko Paalanen for convincing me
that surface acoustic waves in the Quantum Hall effect can not measure the odd
viscosity of the electron gas. I thank Dr. K. Malik for useful suggestions and O.
Regev for a  discussion about Whitehead paradox. This work is supported in part
by  the Israel Academy of Sciences, the DFG, and by the Fund for the Promotion of
Research at the Technion.

\end{document}